\documentclass[amsfonts,floatfix,superscriptaddress,longbibliography,twocolumn]{revtex4-2}

\usepackage{mathbbol}
\usepackage{amsmath,amssymb,bm,amsthm}
\usepackage{epstopdf}
\usepackage{braket}
\usepackage{soul}

\usepackage[utf8x]{inputenc}
\usepackage[usenames,dvipsnames, pdftex]{xcolor}
\usepackage[colorlinks, linkcolor=blue,citecolor=blue, urlcolor=blue]{hyperref}
\usepackage{graphicx}

\usepackage[nodayofweek,level]{datetime}

\begin{document}

\title{Equilibration time in many-body quantum systems}
\author{Tal\'ia L. M. Lezama}
\address{Department of Physics, Ben-Gurion University of the Negev, Beer-Sheva 84105, Israel}
\author{E. Jonathan Torres-Herrera}
\address{Instituto de F\'isica, Benem\'erita Universidad Aut\'onoma de Puebla,
Apt. Postal J-48, Puebla, 72570, Mexico}
\author{Francisco P\'erez-Bernal}
\address{Dep. CC. Integradas y Centro de Estudios Avanzados en F\'isica,
Matem\'aticas y Computaci\'on. Fac. CC. Experimentales, Universidad de Huelva, Huelva 21071, \& Instituto Carlos I de F\'isica Te\'orica y Computacional, Universidad de Granada, Granada 18071, Spain}
\author{Yevgeny Bar Lev}
\address{Department of Physics, Ben-Gurion University of the Negev, Beer-Sheva 84105, Israel}
\author{Lea F. Santos}
\affiliation{Department of Physics, Yeshiva University, New York, New York 10016, USA}

%\date{\mydate}
\begin{abstract}
Isolated many-body quantum systems quenched far from equilibrium can eventually equilibrate, but it is not yet clear how long they take to do so. To answer this question, we use exact numerical methods and analyze the entire evolution, from perturbation to equilibration, of a paradigmatic disordered many-body quantum system in the chaotic regime. We investigate how the equilibration time depends on the system size and observables. We show that if dynamical manifestations of spectral correlations in the form of the correlation hole (``ramp'') are taken into account, the time for equilibration scales exponentially with system size, while if they are neglected, the scaling is better described by a power law with system size, though with an exponent larger than what is expected for diffusive transport.
\end{abstract}
\maketitle

%%%%%%%%%%%%%%%%%%%%%%%%%%%%%%%%%%%%%%%%%%%%%%%%%%%%%
\section{Introduction}
%%%%%%%%%%%%%%%%%%%%%%%%%%%%%%%%%%%%%%%%%%%%%%%%%%%%%

One major question in studies of nonequilibrium dynamics of isolated many-body quantum systems is how long it takes for an experimentally relevant observable to reach equilibrium. By equilibration we mean that after initial transients, the expectation value of the considered quantity exhibits small fluctuations around its infinite-time average, being thus very close to this saturation point for the vast majority of times, and in addition to that, the size of these fluctuations decreases as the system size increases~\cite{Peres:1984,Tasaki:1998,Reimann:2008,Short:2011,Short:2012,Reimann:2012,Zangara:2013,Kiendl2017}. 

The variety of approaches taken to address this question and the lack of agreement among the existing results are worrisome. Several analyses are based on assumptions about the spectrum, observables, and initial conditions, and often provide bounds for the equilibration time. Some suggest that this time should decrease with system size~\cite{Goldstein:2013,Goldstein2015}, others that it should depend weakly on it~\cite{deOliveira:2018}, and others yet that it should increase with it~\cite{Reimann:2008,Reimann:2012,Reimann:2016,Short:2011,Short:2012,Monnai2013,Hetterich2015,Gogolin:rev,Pintos2017}, possibly exponentially~\cite{Goldstein:2013,Malabarba:2014}. Studies aligned with transport behavior~\citep{BarLev:2015,Agarwal:2015,Potter:2015,Vosk:2015,Serbyn_criterion:2015,Znidaric:2016,Luitz_extended:2016,Luitz_anomalous:2016,Bera:2017,Serbyn_thouless:2017,Dymarsky_bound:2018,Luitz:rev,Bertini:rev,Gopalakrishnan:rev}, on the other hand, expect the equilibration time to increase as a power law with system size.

Confronted by so many options, it is worth to step back and try first to identify a general scenario. For this purpose, we focus on many-body quantum systems that are in the chaotic regime and initial states that are far from equilibrium and that have energy expectation values away from the edges of the many-body spectrum. With this choice, we avoid the particularities of integrable models and non-generic initial states.

The largest possible timescale of quantum systems is given by the inverse of their mean-level spacing, the so-called Heisenberg time, which grows linearly with the dimension of the Hilbert space, and thus exponentially with the size of many-body systems. The Heisenberg time is the absolute upper bound for the equilibration time, but do experimental observables take this long to equilibrate? This is the main question addressed by this work. The answer is yes~\cite{Schiulaz:2019} if the dynamical manifestations of spectral correlations, known as correlation hole~\cite{Leviandier:1986,Guhr:1990,Wilkie:1991,Alhassid:1992,Gorin:2002,Torres2017,Torres_Philo:2017,Torres:2018,Schiulaz:2019,Borgonovi_timescales:2019} and sometimes referred to as ``ramp''~\cite{Cotler:2017,Gharibyan:2018,WinerARXIV},  are taken into account. However, as we show here, these manifestations can in practice be neglected for some observables, so that the time for them to reach equilibrium can be defined at a point before the correlation hole. In this case, the equilibration time scales as a power law with system size, a result that is in better agreement with studies of transport behavior~\cite{dAlessio:2016,Gopalakrishnan:rev}.

We use exact numerical methods to study the time evolution of four observables in the chaotic regime of a disordered spin-1/2 Heisenberg chain, which is a general setting for theoretical and experimental studies of the nonequilibrium quantum dynamics of one-dimensional systems with short-range couplings. In addition to examining the scaling of the equilibration time with system size, we also briefly discuss its dependence on the disorder strength. We study two correlation functions of local operators which have been accessed experimentally, namely the spin autocorrelation function and the connected spin-spin correlation function. The former is related to the imbalance used in experiments with cold atoms~\cite{Schreiber:2015} and the latter is measured in experiments with ion traps~\cite{Richerme:2014}. Both are few-body observables and should thus reach thermal equilibrium when the system is in the chaotic regime~\cite{Zelevinsky1996,Borgonovi:2016,dAlessio:2016}. We also study the survival probability, which is the absolute square of the correlation function of the initial state with its evolved counterpart and may be accessible  to experiments with cold atoms~\cite{Gherardini2017,Singh2019}. A semi-analytical expression exists for the evolution of this quantity in the chaotic regime, which provides insights into our analysis~\cite{Schiulaz:2019,Torres:2018}. Our fourth observable is the inverse participation ratio, which describes the spread of the initial state in the many-body Hilbert space and whose logarithm is the participation second-order R\'enyi entropy.

The article is organized as follows. Section~\ref{sec:model} introduces the disordered model, initial conditions, and observables. In Sec.~\ref{sec:hole}, we revisit the concept of the correlation hole, its timescales, and the equilibration time after it. In Sec.~\ref{sec:timescales}, the correlation hole is neglected and a new definition for the equilibration time is proposed. Its dependence on the disorder strength is also provided. Conclusions are presented in  Sec.~\ref{sec:conclusions}. Additional numerical results are presented in the Appendix.

%%%%%%%%%%%%%%%% MODEL %%%%%%%%%%%%%%%%%%%%%
\section{Model, initial states, and observables}
%%%%%%%%%%%%%%%%%%%%%%%%%%%%%%%%%%%%%%%%%%
\label{sec:model}

\subsection{Model}

The disordered spin-1/2 Heisenberg chain that we consider is a representative model of disordered interacting one-dimensional systems and has been extensively used in experimental and theoretical studies of many-body localization~\citep{SantosEscobar:2004,Dukesz2009,Nandkishore:rev,Alet:rev,Abanin:rev,Smith2016}. It is given by the following Hamiltonian,
\begin{equation}
\hat{H}=\sum_{i=1}^{L}J\mathbf{\hat{S}}_{i}\cdot\hat{\mathbf{S}}_{i+1}+h_{i}\hat{S}_{i}^{z},
\label{eq:hamiltonian}
\end{equation}
where  $\mathbf{\hat{S}}_{i}\equiv\left(\hat{S}_{i}^{x},\hat{S}_{i}^{y},\hat{S}_{i}^{z}\right)$ are spin-1/2 operators, $L$ is the system size, periodic boundary conditions are assumed, we set $J=1$ and $h_{i}$ to be independent and uniformly distributed random variables in $[-W,W]$, with $W$ being the onsite disorder strength.
The system conserves the total magnetization in the $z$-direction, $\hat{S}^{z}_{\mathrm{tot}}=\sum_{i}^{L}\hat{S}_{i}^{z}$, and exhibits a transition to the many-body localized phase at a critical disorder strength 
$W_{c}$. The value of $W_{c}$ is still under debate~\citep{Oganesyan:2007,Pal:2010,Berkelbach:2010,Kjall:2014bd,BarLev:2014,Luitz:2015,Devakul2015,Doggen2018,Herviou:2019,Abanin:2021}, some papers estimate that $3<W_{c}<4$  and others that $W_{c}> 4$.

We work in the $\hat{S}^{z}_{\mathrm{tot}}=0$ subspace that has the Hilbert space dimension $D=\binom{L}{L/2}$, and consider finite systems away from the critical region, $0.5\leq W\lesssim 1$, although a short discussion for values of disorder closer to the critical region $1<W<3$ is also provided.

\subsection{Initial states and equilibration}

We use initial states $\ket{\Psi(0)}$ that are product states in the $S^{z}$-basis, $|\phi_n \rangle$, such as $ |\uparrow \downarrow \downarrow \ldots \uparrow \rangle$, which can be experimentally realized. We choose initial states with an energy expectation value $\bra{\Psi(0)}\hat{H}\ket{\Psi(0)}$ far from the edges of the spectrum to guarantee that they fall in the chaotic region of the many-body Hamiltonian and that a given few-body observable $\hat{O}$ reaches thermal equilibrium. In the absence of degeneracies in the spectrum, the infinite time average of $\hat{O}$ expressed in terms of the many-body eigenstates $\hat{H}\ket{\alpha} = E_{\alpha}\ket{\alpha}$ is given by
\begin{equation}
\bar{O} = \sum_{\alpha} |C_{\alpha}|^2 \bra{\alpha} \hat{O} \ket{\alpha} ,
\label{Eq:sat}
\end{equation}
where $C_{\alpha} \equiv \langle \alpha \ket{\Psi(0)} $.    

We use either exact diagonalization or Krylov-space methods to evolve the system in time. All results are averaged over $10^{4}$ samples composed of $0.01D$ initial states and $10^{4}/( 0.01D)$ disorder realizations. The average over samples is denoted by the angle brackets $\langle\cdot\rangle$. We have also verified that fixing a single initial state and using $10^{4}$ disorder realizations leads to equivalent results, though the numerical procedure is less efficient. All statistical errors are accessed using a bootstrap procedure.

\subsection{Observables}
We study the time-evolution of the survival probability and the inverse participation ratio, which are non-local quantities in real space; and two correlation functions of local operators, the spin autocorrelation function and the connected spin-spin correlation function. The last two correlation functions are few-body observables, and therefore are expected to reach thermal equilibrium in the chaotic limit of realistic systems~\cite{dAlessio:2016}. 

The survival probability is defined as
\begin{equation}
P_{S}(t)=\left|\left<\Psi(0)|\Psi(t)\right>\right|^{2}.
\label{eq:PS}
\end{equation}
Taking the mean gives
\begin{equation}
\langle P_{S}(t) \rangle = \langle \sum_{\alpha \neq \beta}  |C_{\alpha}|^2 |C_{\beta}|^2 e^{- i (E_{\alpha} - E_{\beta}) t} \rangle +  \langle \sum_{\alpha} |C_{\alpha}|^4 \rangle,
\end{equation}
which is related to the spectral form factor, 
$\langle \sum_{\alpha \neq \beta}  e^{- i (E_{\alpha} - E_{\beta}) t} \rangle$.
While $\langle P_{S}(t) \rangle$ is a true dynamical quantity, which depends on the initial state through the components $C_{\alpha}$, the  spectral form factor involves only the eigenvalues of the Hamiltonian and is used to study the manifestation of level statistics in the time domain~\cite{MehtaBook}. Filter functions are sometimes added to it~\cite{Prosen_challenges:2020,WinerARXIV}, but they do not come from the quench dynamics, as the coefficients $C_{\alpha}$  in $\langle P_{S}(t) \rangle$. The survival probability is widely used in studies of nonequilibrium quantum dynamics and quantum speed limit. Both the survival probability~\cite{Schiulaz:2019} and the spectral form factor~\cite{Prosen_challenges:2020} have been used in the analysis of the many-body localization transition and exhibit a robust correlation hole in the chaotic regime, which fades away as the system approaches the many-body localized phase~\cite{Torres2017,Torres_Philo:2017,Torres:2018,Schiulaz:2019,Prosen_challenges:2020}.

The other non-local quantity that we consider is the inverse participation ratio, 
\begin{equation}
\text{I}_\text{PR}(t)=\sum_{n} \left|\left\langle \phi_n|\Psi(t)\right\rangle\right|^4,
\label{eq:IPR}
\end{equation}
which measures the spreading in time of the initial state $|\Psi(0)\rangle$ over the many-body Hilbert space; the symbol $|\phi_n\rangle$ denotes a state of the computational basis. Its logarithm, $- \ln \text{I}_\text{PR}(t)$, corresponds to the second-order R\'enyi entropy. The minimum of the inverse participation ratio indicates the full spread of the initial state in its energy shell.

The spin autocorrelation function measures the proximity of the spin configuration in the $z$-direction at a time $t$ to the initial spin configuration,
\begin{equation}
\mathcal{I}(t)=\frac{4}{L}\sum_{i=1}^{L}\left<\Psi_{0}\right|\hat{S}_{i}^{z}\left(0\right)\hat{S}_{i}^{z}\left(t\right)\left|\Psi_{0}\right>,\label{eq:I}
\end{equation}
where $\hat{S}_{i}^{z}\left(t\right)=e^{iHt}\hat{S}_{i}^{z}(0)e^{-iHt}$. For the N\'eel initial state, $|\uparrow \downarrow  \uparrow \downarrow \uparrow \downarrow \ldots \rangle $, it reduces to the density imbalance between
even and odd sites that is measured in experiments with cold atoms~\citep{Schreiber:2015}.

The connected spin-spin correlation function is defined as
\begin{eqnarray}
C(t)  &=&\frac{4}{L}\sum_{i=1}\left[\left<\Psi(t)\right|\hat{S}_{i}^{z}\hat{S}_{i+1}^{z}\left|\Psi(t)\right> \right. \nonumber \\
 &-& \left. \left<\Psi(t)\right|\hat{S}_{i}^{z}\left|\Psi(t)\right>\left<\Psi(t)\right|\hat{S}_{i+1}^{z}\left|\Psi(t)\right>\right]
\end{eqnarray}
and has been measured in experiments with ion traps~\citep{Richerme:2014}.

%%%%%%%%%%%%%%%%%%%%%%%%%%%%%%%%%%%%%%%%%%%%%%%%%%%%%
%%%%%%%%%%%%%%% THERMALIZATION TIME AFTER THE HOLE %%%%%%%%%%%%%%%
\section{Equilibration after the correlation hole}
%%%%%%%%%%%%%%%%%%%%%%%%%%%%%%%%%%%%%%%%%%%%%%%%%%%%%
\label{sec:hole}

In this section we analyze the appearance of the correlation hole for the four observables introduced above and discuss the use of the time after the interval of the hole as a definition of the equilibration time. How the timescale for the correlation hole depends on the system size $L$ and on the disorder strength $h$ \cite{Schiulaz:2019} and how its depth depends on $h$ ~\cite{Torres2017,Torres_Philo:2017,Torres:2018} were studied before for the survival probability and the spin autocorrelation function. Here, we investigate how the depth of the correlation hole depends on the system size and whether it survives in the thermodynamic limit for the four quantities considered.

A semi-analytical expression for the entire evolution of the average of the survival probability was derived for realistic chaotic many-body quantum systems with short-range interactions~\cite{Torres:2018,Schiulaz:2019}, as the one described by Eq.~(\ref{eq:hamiltonian}), and it is given by
\begin{equation}
\left< P_S(t)\right> =\frac{1- \left< \overline{P_S} \right> }{ (D-1)}\left[ D b_1^2(\Gamma t) -b_2\left(\frac{\Gamma t}{\sqrt{2\pi}D} \right)\right] + \left< \overline{P_S} \right>,
\label{Eq:PsGOE} 
\end{equation}
where 
\begin{equation}
\Gamma^2 = \langle \Psi(0) |\hat{H}^2| \Psi(0) \rangle - \langle \Psi(0) |\hat{H}| \Psi(0) \rangle^2
\end{equation}
is the squared width of the energy distribution of the initial state, $\left< \overline{P_S} \right>$ is the mean of the infinite-time average of $P_S(t)$,
\begin{equation}
b_1^2(\Gamma t) \!=\! \frac{ e^{-\Gamma^2t^2}}{4{\cal N}^2} \!  \left|\textrm{erf}\left(\frac{E_\textrm{max}+it\Gamma^2}{\sqrt{2}\Gamma}\right) \!\! - \! \textrm{erf}\left(\frac{E_\textrm{min}+it\Gamma^2}{\sqrt{2}\Gamma}\right)\right|^2 \!\!,
\end{equation}
$\mathcal{N}$ is a normalization constant, $\textrm{erf}$ is the error function, $E_\textrm{max}$ and $E_\textrm{min}$ are the largest and smallest eigenvalues of $\hat{H}$, respectively, and
\begin{equation}
b_2(t) =\begin{cases}
1-2t + t \ln(2 t+1) & t\leq1 \\
t \ln \left( \dfrac{2 t+1}{2 t -1} \right) -1 & t>1,
\end{cases}
\label{eq:b2}
\end{equation}
is the two-level form factor. 

\begin{figure*}[htb]
\centering{}
\includegraphics[width=1\textwidth]{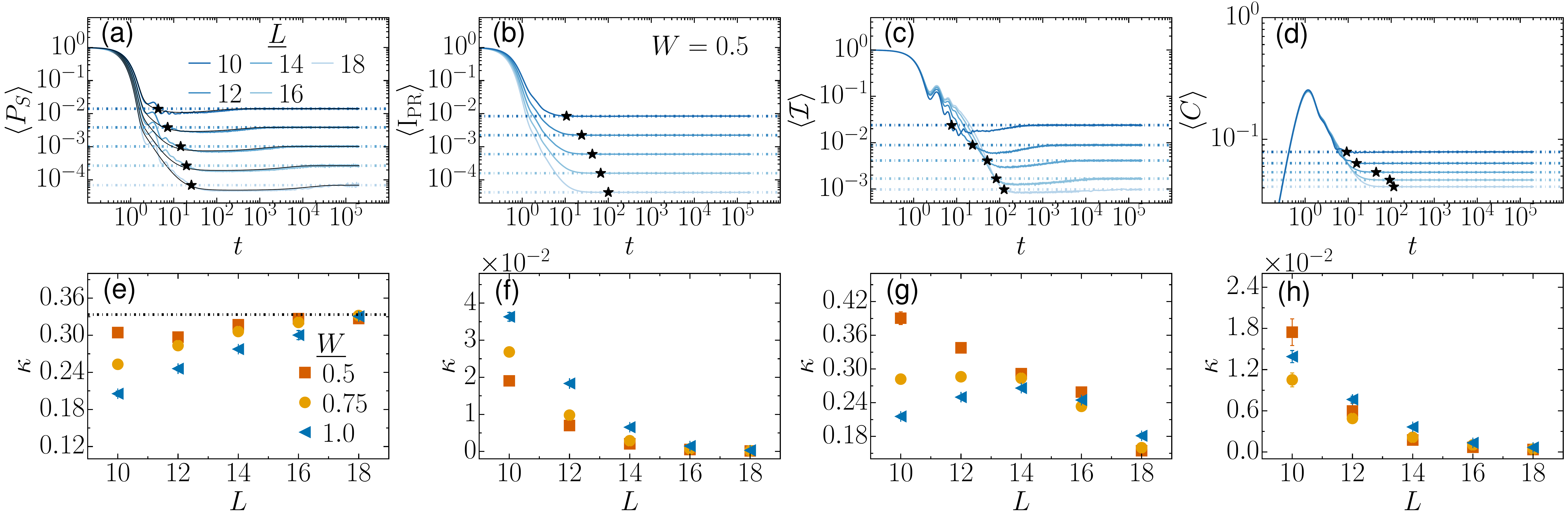} 
\caption{\emph{Upper panels}: Time evolution of the mean (a) survival probability $\langle P_{S}(t) \rangle$, (b) inverse participation ratio $\langle I_{\text{PR}}(t) \rangle$, (c) spin autocorrelation function $\langle \mathcal{I}(t) \rangle$, and (d) connected spin-spin correlation function $\langle C(t)\rangle$ for different system sizes, as indicated in panel (a), and for disorder strength $W=0.5$, as shown in panel (b). The horizontal dotted-dashed lines mark the asymptotic values and the stars indicate the crossing time $t^*$. In Fig.~\ref{fig:hole}~(a), the numerical data overlap with the curves for the semi-analytical expression in Eq.~(\ref{Eq:PsGOE}).   \emph{Lower panels}:  Relative depth of the correlation hole, $\kappa$, as a function of $L$, for (e) $\langle P_{S}(t) \rangle$, (f) $\langle I_{\text{PR}}(t) \rangle$, (g) $\langle \mathcal{I}(t) \rangle$, and (h) $\langle C(t)\rangle$ for three values of the disorder strength in the chaotic regime, as indicated in Fig.~\ref{fig:hole}~(e). The horizontal dashed line in panel (e) corresponds to the value $\kappa=1/3$ obtained for GOE random matrices. Most error bars in panels (e)-(h) are smaller than the symbols.}
\label{fig:hole}
\end{figure*}

The decay of the survival probability is controlled by $b_1^2(\Gamma t)$. This function and the $b_2$ function meet at the time $t_m \propto D^{2/3}/\Gamma$, where the survival probability reaches its minimum value of  $\sim 2/D$.
 After this point, the evolution is described entirely by the $b_2$ function, which brings $\left< P_S(t)\right>$ from its minimum up to the saturation value, which is $\sim 3/D$. Saturation happens at the Heisenberg time $t_{\text{H}} \propto D/\Gamma$. 

The time interval governed by the $b_2$ function, where $\left< P_S(t)\right> < \left< \overline{P_S} \right>$, is known as correlation hole~\cite{Leviandier:1986,Guhr:1990,Wilkie:1991,Alhassid:1992,Gorin:2002,Torres_Philo:2017,Torres:2018,Schiulaz:2019} or ``ramp''~\cite{Cotler:2017,Gharibyan:2018,WinerARXIV}. This is a dynamical manifestation of spectral correlations, and, therefore, a dynamical signature of quantum chaos. The point in time where the ramp starts, $t_m$, has been referred to as Thouless time~\cite{Schiulaz:2019,Prosen_challenges:2020}. It coincides with the time where the inverse participation ratio reaches its minimum value~\cite{Schiulaz:2019}, indicating that $t_m$ is the time that it takes for the initial state to maximally spread over the many-body Hilbert space and acquire weight over the unperturbed many-body states $|\phi_n \rangle$ in its microcanonical energy shell given by the width $\Gamma$.

The evolution of the mean survival probability for the spin model (\ref{eq:hamiltonian}) is shown in Fig.~\ref{fig:hole}~(a) for different system sizes. There is an excellent agreement between the numerical results and expression (\ref{Eq:PsGOE}) when $W=0.5$, which corresponds to the deep chaotic regime. The correlation hole is evident in all the curves, after a sufficient number of samples is used for the averages~\cite{Schiulaz2020}, and it does not fade away as $L$ increases. This means that the complete equilibration of this quantity takes place only after the hole ends at the Heisenberg time $t_{\text{H}}$.  Since this time is exponentially long in the system size $L$, we use exact diagonalization to resolve the entire dynamics, which limits the accessible systems sizes to $L=18$.

A correlation hole is also visible for the spin autocorrelation function, as depicted in Fig.~\ref{fig:hole}~(c), suggesting that for sufficiently small system sizes, where it develops for times that are not exceedingly long and reaches minimum values that are not too small, the hole might be experimentally detected.

In contrast to the survival probability and the spin autocorrelation function, the effects of the spectral correlations in the evolution of $\langle \text{I}_\text{PR}(t)\rangle$ [Fig.~\ref{fig:hole}~(b)] and of $\langle C(t) \rangle$ [Fig.~\ref{fig:hole}~(d)] are minor and the correlation hole is hardly discernible. Furthermore, the analysis in Figs.~\ref{fig:hole}~(e)-(h) of the relative depth of the correlation hole~\citep{Torres2017,Torres_Philo:2017,Torres_signatures:2019},
\begin{equation}
\kappa=\frac{ \langle \overline{ O} \rangle- \langle O \rangle_{\text{min}}}{ \langle \overline{ O} \rangle},
\label{Eq:kappa}
\end{equation}
where $\langle O \rangle_{\text{min}}$ stands for the value of a given observable $\hat{O}$ at the minimum of the correlation hole, indicates that, contrary to what happens for $\left< P_S(t)\right>$, the hole for $\langle \text{I}_\text{PR}(t)\rangle$, $\langle \mathcal{I}(t) \rangle $, and $\langle C(t) \rangle$ gets smaller as the system size increases. This motivates an alternative definition of the equilibration time for these three quantities, which neglects the correlation hole, as done below in Sec.~\ref{sec:timescales}. 

When the survival probability is evolved using full random matrices taken from a Gaussian orthogonal ensemble (GOE), it is known analytically that $\left< \overline{P_S} \right>\sim 3/D$ and $\langle P_S \rangle_{\text{min}}\sim 2/D$ \cite{Alhassid:1992,Schiulaz:2019}, which yields $\kappa=1/3$. In Fig.~\ref{fig:hole}~(e), we show $\kappa$ for the survival probability of the spin model as a function of system size for different disorder strengths in the chaotic regime. The relative depth clearly converges to $\kappa=1/3$, which is indicated with a horizontal dashed line. The correlation hole is therefore a robust property of the survival probability.

\begin{figure*}[tb!]
\centering{}
\includegraphics[width=1\textwidth]{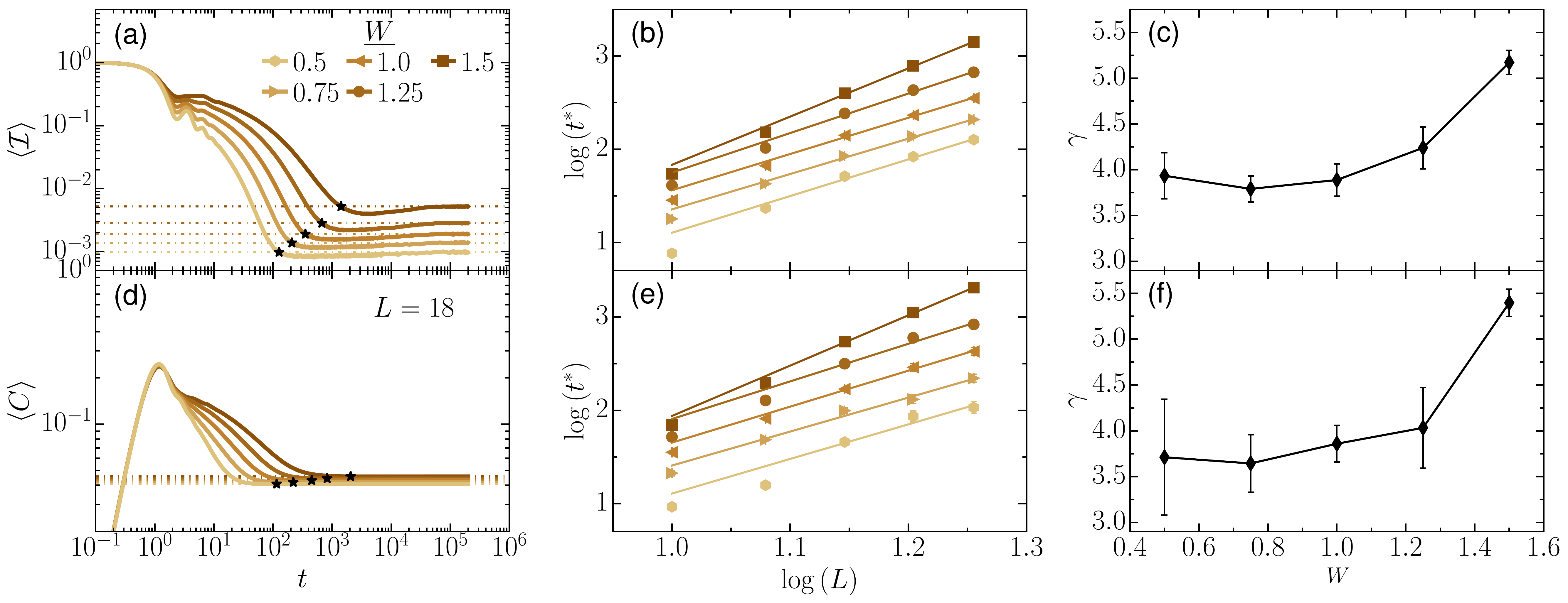} 
\caption{\emph{Upper [Lower] panels}: Analysis of the crossing time for the spin autocorrelation function $\langle \mathcal{I}(t) \rangle $ [connected spin-spin autocorrelation function $\langle C(t) \rangle $].
(a) [(d)] Evolution of the mean spin autocorrelation function [connected spin-spin autocorrelation function]  for $L=18$. Each curve corresponds to a different disorder strength $W$, as indicated. The horizontal dotted-dashed lines represent the asymptotic value of the observable and the crossing time $t^*$ is marked with a star. (b) [(e)] Scaling of the crossing time $t^{*}$ with $L$. Symbols are the data and solid lines give power-law fits to $t^{*}\propto L^{\gamma}$. Error bars indicate the standard deviation of the bootstrap procedure. (c) [(f)] Exponent~$\gamma$ extracted from the power-law fit to the data points in (b) [(e)] as a function of $W$. Error bars indicate the standard deviation on the fitted exponent.}
\label{fig:star}
\end{figure*}

Contrary to the survival probability, the relative depth $\kappa$ for $\langle \text{I}_\text{PR}(t)\rangle$ [Fig.~\ref{fig:hole}~(f)], $\langle \mathcal{I}(t) \rangle$ [Fig.~\ref{fig:hole}~(g)], and $\langle C(t) \rangle$ [Fig.~\ref{fig:hole}~(h)]  decays with $L$. In the case of the inverse participation ratio and the connected spin-spin correlation function, the decay is exponential, while for the spin autocorrelation function, the results are more subtle and make apparent the danger of finite-size effects. While for $W=0.5$, $\kappa$ decreases monotonically  with $L$, for $W=0.75$ and $W=1$, $\kappa$ increases for small values of $L$ and the decay becomes clear only for $L > 14$.

One sees that the dynamical behavior of the four quantities considered transcends any simple categorization in terms of locality or non-locality in real space. One might be tempted to associate the visible onset of the correlation hole, at least for the relatively small system sizes that we study, with non-locality in time. However, to confirm this speculation, quantities other than the spin autocorrelation function and the survival probability, which is also an autocorrelation function, need to be investigated.

%%%%%%%%%%%%%%%%%%%%%%%%%%%%%%%%%%%%%%%%%%%%%%%%%%%%%%
%%%%%%%%%%%%%%% THERMALIZATION TIME BEFORE THE HOLE %%%%%%%%%%%%%%%
\section{Equilibration before the correlation hole}
%%%%%%%%%%%%%%%%%%%%%%%%%%%%%%%%%%%%%%%%%%%%%%%%%%%%%%
\label{sec:timescales}

Ignoring the correlation hole, we can define the equilibration time as the point where $\langle \hat{O}(t) \rangle$ first crosses the infinite-time average $\langle \overline{O} \rangle$. We denote this time by $t^*$, which is clearly much smaller than the Heisenberg time, $t^{*} \ll t_H$. These crossing points are marked with stars in Figs.~\ref{fig:hole}~(a)-(d). The fact that the correlation hole exists for a finite-size system, even if this is minor, indicates that $t^{*}$ is well defined, because finding its value consists of finding a crossing point. This is to be contrasted with the determination of the Heisenberg time $t_{\text{H}}$. Since the $b_2(t)$ function, which controls the evolution in the interval of the correlation hole, follows a power-law behavior at long times, finding $t_{\text{H}}$ relies on an arbitrary threshold $\delta$ between the observable and its infinite-time average, $|\langle O(t_{\text{H}})\rangle - \overline{O}|/\overline{O}  \sim \delta$, as discussed in~\cite{Schiulaz:2019}.

\subsection{Weak disorder region: $\pmb{0.5\leq W \leq1}$}
\label{sec:timescalesA}

For sufficiently weak disorder, $W=0.5$~\cite{noteInt}, we can make use of the semi-analytical expression in Eq.~(\ref{Eq:PsGOE}) to estimate the dependence of $t^{*}$ on system size. At long times, disregarding the correlation hole, the decay of  $\left< P_S (t) \right>$ is given by~\cite{Tavora2016,Tavora2017}
\begin{equation}
\left< P_S (t\gg \Gamma^{-1}) \right>_{decay} \simeq \frac{  e^{-E_{\text{max}}^2/\Gamma^2} + e^{-E_{\text{min}}^2/\Gamma^2} }{2 \pi {\cal N}^2 \Gamma^2 t^2} .
\end{equation}
Using that $E_{\text{min}}$, $E_{\text{max}}$ and $\Gamma^2$ are extensive, namely proportional to the size of the system, and that the survival probability saturates at $\left< \overline{P_S} \right> \sim 3/D$, we find that $t^* \propto \exp(0.22L)$, which agrees very well with the crossing time obtained numerically for $L=10,12,14,16,18,20,22$.
Knowing the saturation value of the survival probability, we can evolve $|\Psi(t) \rangle$ up to a vicinity of  $t^*$ only, which is a major saving compared to the evolution up to $t_\text{H}$. This allows us to use Krylov-space methods for the time evolution of the survival probability and access to system sizes $L>18$. For the seven system sizes considered, we verified that the exponential scaling of $t^*$ with $L$ is indeed better than a power-law scaling.

While an analytical expression is not available for the time-dependence of the inverse participation ratio, in the weak disorder regime, its saturation value is known analytically, $\left< \overline{I_\text{PR}} \right> \sim 2/D$ \cite{Torres2020b}, so we can also obtain $t^*$ for system sizes $L>18$. To do that, we apply a Savitzky-Golay filter to smooth the curves for $\left< I_\text{PR} (t) \right>$ and then extract the crossing time~\cite{footSmooth}. For this quantity, which is non-local in real space just as the survival probability, we find that a power-law scaling of $t^*$ with $L$ actually works better than an exponential scaling. This makes us suspect that the exponential dependence of the crossing time with system size found for the survival probability may be related to the prevalence of the correlation hole, a conjecture that is further reinforced by the next results.

For the two few-body observables, we do not have analytical results for the saturation values, hence our analysis is restricted to five system sizes. In Fig.~\ref{fig:star}~(a) and Fig.~\ref{fig:star}~(d), we fix the system size at $L=18$ and mark the crossing time for the curves of $\langle \mathcal{I}(t) \rangle$ and $\langle C(t) \rangle$ obtained for different disorder strengths. In Fig.~\ref{fig:star}~(b)  and Fig.~\ref{fig:star}~(e), we present the scaling analysis for both quantities and those values of $W$. We find that, similarly to the inverse participation ratio, the dependence of $t^*$ with $L$ is better fitted with a power law than an exponential, that is
\begin{equation}
t^{*}\propto L^{\gamma}\quad\mathrm{with}\quad\gamma>3.
\label{eq:gamma}
\end{equation}
The exact value of the exponent $\gamma$ varies with the disorder strength, as shown in Fig.~\ref{fig:star}~(c) and Fig.~\ref{fig:star}~(f). For disorder strength $0.5\leq W \leq 1$, we find that on average $\gamma\approx3.8\pm 0.3$. The value of the exponent is somewhat close to that argued in~\cite{Dymarsky_bound:2018}, but larger than the value which typically appears in studies of transport behavior~\cite{dAlessio:2016,Gopalakrishnan:rev}. In particular, in~\cite{Dymarsky_bound:2018}, it was argued that matrix elements of local operators in the energy basis of chaotic Hamiltonians remain correlated down to frequencies parametrically lower (corresponding to parametrically larger times) than those expected from the diffusive scaling, beyond which true random-matrix behavior occurs, while other measures of thermalization seemed otherwise fulfilled. The observation that these elements remain correlated down to such low frequencies was further tested numerically in~\cite{Richter:2020}, although the scaling of a critical frequency with system size could not be obtained.

We stress that the system has to be evolved for very long times to obtain the time for the saturation of the dynamics, which limits the system sizes that can be explored. This is particularly problematic for the few-body observables, where only 5 system sizes were considered. Thus, any conclusion regarding the scaling analysis over such small range of points should be taken with reservation.  

We expect to obtain results analogous to those in Sec.~\ref{sec:hole} and Sec.~\ref{sec:timescalesA} for chaotic clean models, since the correlation hole is, of course, present there as well. However, the procedure to extract the equilibration time requires sufficiently smooth curves, which is more difficult to achieve in the absence of disorder average. To reveal the correlation hole in clean models, one needs to resort to averages over initial states and running averages. In the Appendix~\ref{App1}, we show the evolution of the survival probability and the spin autocorrelation function for a model without onsite disorder, so that the reader can see that the behaviors are similar to those in Fig.~1, but the curves are much less smooth.

\subsection{Near critical region: $\pmb{1 < W <W_c}$}

Figure~\ref{fig:star} shows the crossing time $t^*$ of the few-body observables computed for disorder strengths slightly larger than the coupling parameter, $W \gtrsim J$. The power-law fit is still better than the exponential one, although $\gamma$ [Eq. (\ref{eq:gamma})] for $W=1.5$ is larger than 5, as seen in Fig.~\ref{fig:star}~(c) and Fig.~\ref{fig:star}~(f). The scaling analysis with $L$ of the crossing time for $W>1$ is more difficult, because in this region, the use of random matrices for guidance is no longer justified and longer propagation times are typically required to obtain equilibration. We leave open the question of whether the scaling with system size remains power law with even larger exponents or becomes exponential as the critical point is approached (see related discussions in~\cite{Gopalakrishnan:rev,Serbyn_thouless:2017}). In the following, we analyze how the crossing time $t^*$ depends on the disorder strength for a fixed $L$. 

\begin{figure}[tb!]
\centering{}
\includegraphics[width=1\columnwidth]{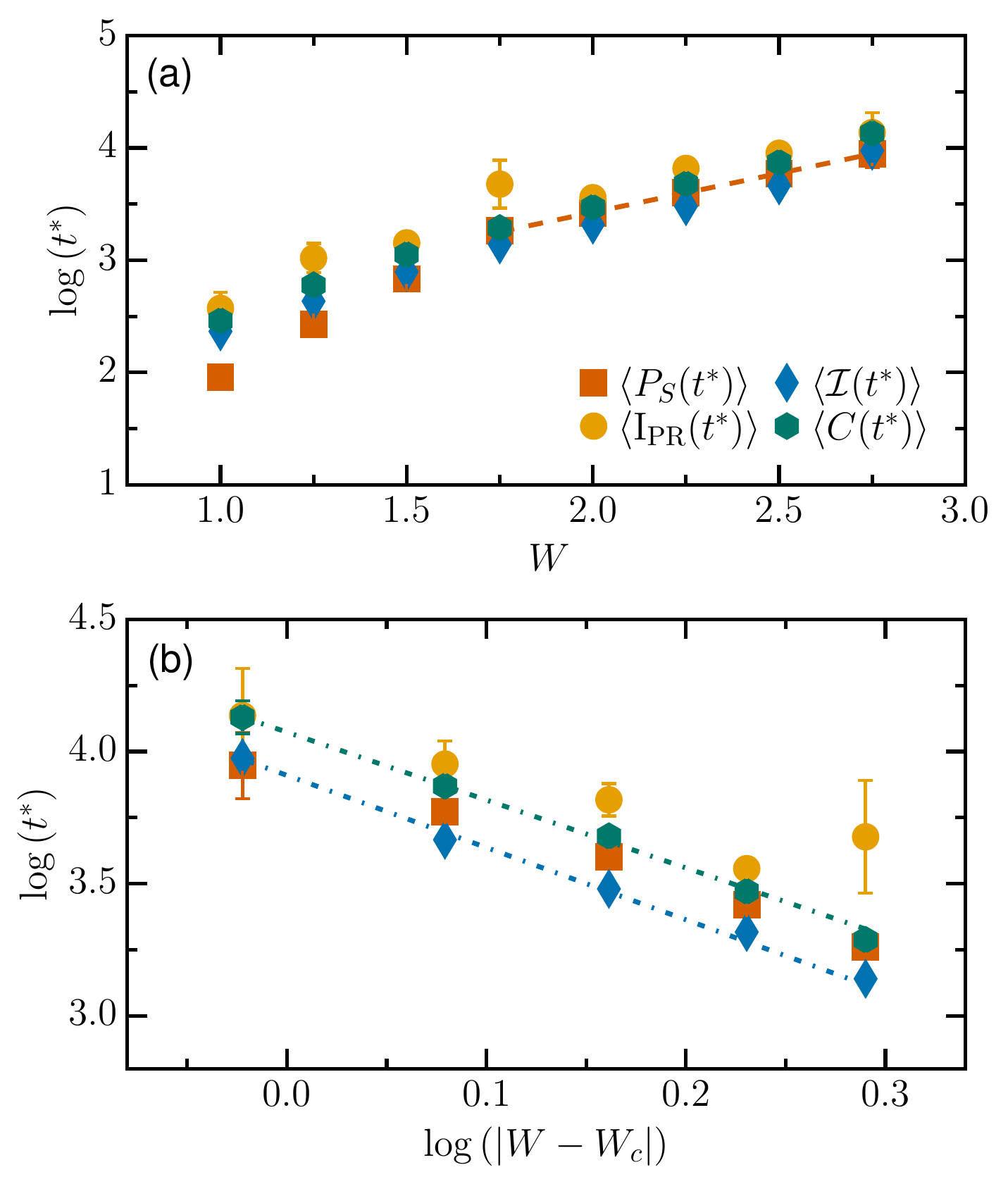} 
\caption{Dependence of the crossing time $t^*$ on (a) [(b)] the disorder strength $W$ [the difference $|W-W_c|$; with $W_c=3.7$] for  the survival probability $\langle P_{S}(t) \rangle$, inverse participation ratio $\langle I_{\text{PR}}(t) \rangle$,  spin autocorrelation function $\langle \mathcal{I}(t) \rangle$, and  the connected spin-spin correlation function $\langle C(t)\rangle$ for system size $L=16$. The symbols representing each quantity are shown in panel (a). The dashed [dotted-dashed] lines in (a) [(b)] correspond to  exponential [power-law] fits obtained with the points for $W \geq 1.75$. Error bars indicate the standard deviation of the bootstrap procedure.
}
\label{fig:disorder}
\end{figure}

It was shown that the time for the minimum of the correlation hole, $t_m$, for the survival probability and for the spin autocorrelation function grows exponentially as the disorder strength increases~\cite{Schiulaz:2019}, which was later confirmed for the spectral form factor~\cite{Prosen_challenges:2020}. It is thus relevant to examine the dependence of $t^*$ on $W$, although its behavior is not conclusive, as can be seen from Fig.~\ref{fig:disorder}. Considering disorder strengths $W \in [1,3]$, the time $t^*$ is best fitted with a stretched exponential, but if we consider disorder strengths closer to the critical point, $W \geq 1.75$, we see that either an exponential dependence $t^*\propto\exp(W/W')$ or a critical form $t^*\propto\left|W-W_c\right|^{-\beta}$ with an exponent $\beta\approx 2.6\pm 0.08$ describe the data reasonably well for all the quantities, except for the inverse participation ratio, which exhibits strong fluctuations in its corresponding crossing times. The quality of the fits varies depending on the quantity:  the survival probability is better described by an exponential form [Fig.~\ref{fig:disorder}~(a)], while the few-body observables seem to have critical scaling [Fig.~\ref{fig:disorder}~(b)]. With respect to Fig.~\ref{fig:disorder}~(b), we did not see a qualitative change in our conclusions when varying $W_c$ from 3.5 to 6.

\subsection{Alternative definition of the equilibration time}

The two definitions for the equilibration time proposed in this work lead to a time that increases with the system size. The observation that the equilibration time increases with system size can be intuitively understood as the time it takes for an initial excitation to visit the full system, a situation that is expected to occur in systems with extensive conserved quantities where there is transport, such as the Hamiltonian with short-range couplings in Eq.~(\ref{eq:hamiltonian}). 

The experimental confirmation of our results depends on several factors, such as the ability to reach very long coherence times, the accuracy of the measurements, and the quantity measured. Consider, for example, the inverse participation ratio in Fig.~\ref{fig:hole}~(b), whose minimum value decreases with increasing system size. If the experimental precision would be limited to $10^{-2}$ one might conclude that the equilibration time decreases as $L$ increases. If, however, the experiment would directly measure the entropies, $\ln [\text{I}_\text{PR}(t)] $ or $ -  \sum_{n} \left|\left\langle \phi_n|\Psi(t)\right\rangle\right|^2 \ln [\left|\left\langle \phi_n|\Psi(t)\right\rangle\right|^2 ] $, that limiting resolution would be circumvented. In practice, it is therefore worthwhile to keep in mind possible discrepant conclusions between theory and experiment due to experimental limitations.

%%%%%%%%%%%%%%%%%%%%%%%%%%%%%%%%%%%%%%%%%%%%%%%%%%%%%%
%%%%%%%%%%%%%%%%%%%%%%%% CONCLUSION %%%%%%%%%%%%%%%%%%%%%%
\section{Conclusions}
%%%%%%%%%%%%%%%%%%%%%%%%%%%%%%%%%%%%%%%%%%%%%%%%%%%%%%
\label{sec:conclusions}

We investigated how the equilibration time of the disordered spin-1/2 Heisenberg chain depends on the system size $L$ for four different observables. For chaotic systems and few-body observables, this time can also be identified with the thermalization time. If the correlation hole is taken into account when defining the equilibration time, then the latter coincides with the Heisenberg time and thus grows exponentially with $L$. This is what happens for the survival probability, where the correlation hole persists in the thermodynamic limit. However, for the inverse participation ratio, the spin autocorrelation function, and the connected spin-spin correlation function, the correlation hole fades away as $L$ increases, which justifies neglecting it. In this case, we defined the equilibration time as the point where the evolution of the observables first crosses their infinite-time averages. The dependence of this crossing time on system size is best described by a power law. 

Chaotic systems with static Hamiltonians conserve at least the total energy, have diffusive energy transport and their equilibration time, namely the time it takes for a non-uniformity in the energy density to spread across the system, is expected to be bounded from below by $L^2$. In the particular case of disordered chaotic systems, transport is presumably subdiffusive   \cite{BarLev:2014,BarLev:2015,Agarwal:2015,Luitz_extended:2016,Luitz:rev}, in which case their equilibration time is bounded from below by $L^\gamma$ with $\gamma>2$, although for very weak disorder, $\gamma$ should eventually approach the lower bound $\gamma\approx 2$ corresponding to diffusive transport. Interestingly, even for the lowest disorder  that we study, $W\approx0.5$, the equilibration (crossing) time for the few-body observables considered scales as $L^\gamma$ with $\gamma > 3$, so it is parametrically larger than the time it takes to make the energy density homogeneous. In future studies, we plan to explore in more detail in which sense the system seems to remain out of equilibrium for times $L^2 < t < L^3$, where the energy density has already spread out, but other few-body observables, such as the spin autocorrelation function and the connected spin-spin correlation function, have not yet equilibrated.

We have provided a brief analysis of the dependence of the crossing time $t^*$ on the disorder strength close to the critical point, but it is hard to discern between an exponentially growing $t^*$ with $W$ and a critical behavior $t^*\propto\left|W-W_c\right|^{-\beta}$. The survival probability seems to be better described by the former, while the few-body observables appear to show the critical scaling with an exponent $\beta\approx 2.6$. 

In summary, by analyzing the entire evolution of physical observables up to equilibration in a paradigmatic many-body quantum system, we were able to identify their equilibration time without any assumptions or approximations. Taking the correlation hole into account, the equilibration time increases exponentially with the system size. Disregarding the correlation hole, the equilibration time for the few-body observables considered and the inverse participation ratio grows as a power law with the system size, although still exponentially for the survival probability. We leave it as an open question for future studies to determine whether this apparent difference in the scaling is related to the vanishing of the correlation hole in the thermodynamic limit.

%%%%%%%%%%%%%%%%%%%%%%%%%%%%%%%%%%%%%%%%%%%%%%%%%%%%%%
\begin{acknowledgments}
This research was supported by a grant from the United States-Israel Binational Foundation (BSF, Grant No. 2019644), Jerusalem, Israel, and the United States National Science Foundation (NSF, Grant No. DMR-1936006), and by the Israel Science Foundation (grants No.  527/19 and 218/19). T.L.M.L. acknowledges funding from the Kreitman fellowship. E.J.T.-H. is grateful to LNS-BUAP for their supercomputing facility. Computing resources supporting this work were provided by the CEAFMC and Universidad de Huelva High Performance Computer (HPC$@$UHU) located in the Campus Universitario el Carmen and funded by FEDER/MINECO project UNHU-15CE-2848. F.P.B. thanks Spanish National Research, Development, and Innovation plan (RDI plan) under the project PID2019-104002GB-C21 and the Consejer\'{\i}a de Conocimiento, Investigaci\'on y Universidad, Junta de Andaluc\'{\i}a and European Regional Development Fund (ERDF), refs. SOMM17/6105/UGR and UHU-1262561.
\end{acknowledgments}

%%%%%%%%%%%%%%%%%%%%%%%%%%%%%%%%%%%%%%%%%%%%%%%%%%%%%%
\appendix
\section{Clean model}
\label{App1}

We consider here a one-dimensional XXZ chain with periodic boundary conditions, but in contrast to Eq.~(\ref{eq:hamiltonian}), it does not have random onsite disorder and contains instead a single defect (impurity) on site $L/2$. The Hamiltonian is given by
\begin{equation}
\hat{H}=   h_{L/2}\hat{S}_{L/2}^{z} + J \sum_{i=1}^{L} \left(\hat{S}^{x}_{i}\hat{S}^{x}_{i+1}+\hat{S}^{y}_{i}\hat{S}^{y}_{i+1} + \Delta\hat{S}_{i}^{z}\hat{S}_{i+1}^{z} \right).
\label{eq:hamiltonianClean}
\end{equation}
We set the coupling constant to $J=1$, the anisotropy parameter to $\Delta=1.2$, and  $h_{L/2}=1.0$, which guarantees that the system is chaotic~\cite{Santos:2004,Santos2011}. We add two small defects, $h_{1}\hat{S}_{1}^{z}$ and $h_{L}\hat{S}_{L}^{z}$, where $h_{1,L}$ are small numbers uniformly distributed in $[-0.1,0.1]$, to break symmetries and to use the averages over realizations to reduce finite-size effects. As in the main text, we work in the $\hat{S}^{z}_{\mathrm{tot}}=0$ subspace. 

\begin{figure}[t!]
\centering{}
\includegraphics[width=1\columnwidth]{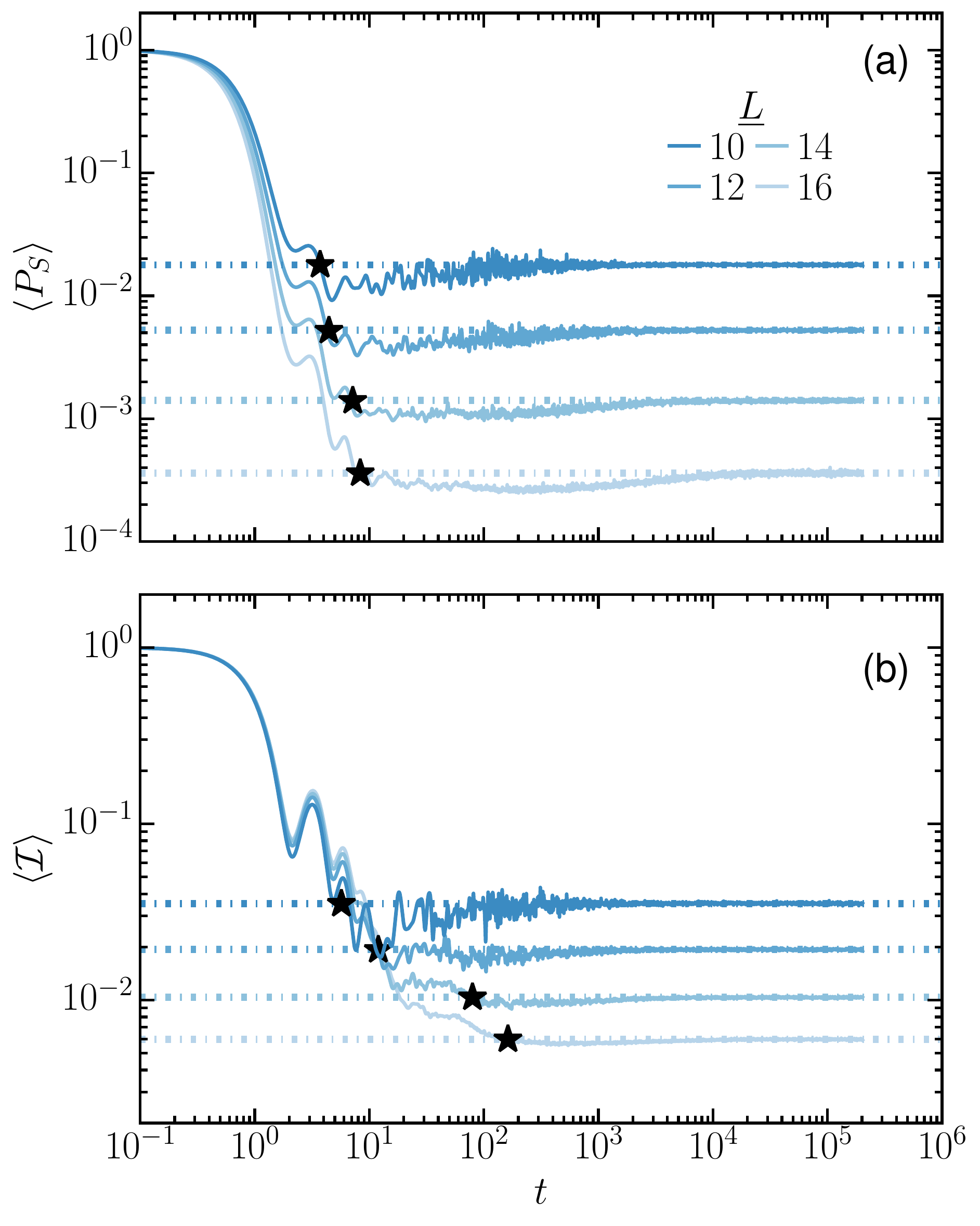} 
\caption{Time evolution of the mean (a) survival probability $\langle P_{S}(t) \rangle$ and (b) spin autocorrelation function  for different system sizes, as indicated in panel (a), of the clean chaotic model in Eq.~(\ref{eq:hamiltonianClean}). The horizontal dotted-dashed lines mark the asymptotic values and the stars indicate the crossing time~$t^*$.
}
\label{fig:clean}
\end{figure}

In Fig.~\ref{fig:clean}, we show the time evolution of the survival probability [Fig.~\ref{fig:clean}~(a)] and the spin autocorrelation function [Fig.~\ref{fig:clean}~(b)]. All results are averaged over $10^{4}$ samples composed of $0.01D$ random product states in the $S^{z}$-basis and $10^{4}/( 0.01D)$ realizations for $h_{1,L}$. Figure~\ref{fig:clean} exhibits features very similar to those in Fig.~\ref{fig:hole}~(a) and Fig.~\ref{fig:hole}~(c), including the appearance of the correlation hole, but the curves are now visibly much less smooth than the corresponding curves in Fig.~\ref{fig:hole}.

%\vspace{1cm}

\newpage
%%%%%%%%%%%%%%%%%%%%%%%%%%%%%%%%%%%%%%%%%%%%%%%%%%%%%%
%\bibliographystyle{apsrev4-1}
\bibliography{references}

\begin{thebibliography}{79}
\bibitem [{not()}]{noteInt}%
  \BibitemOpen
  \href@noop {} {}\bibinfo {note} {Since for $W=0$ the system is integrable,
  the disorder cannot be too small.}\BibitemShut {Stop}%
\bibitem [{foo()}]{footSmooth}%
  \BibitemOpen
  \href@noop {} {}\bibinfo {note} {The statistical errors of this procedure
  were accounted for using a bootstrapping procedure with $1000$ bootstrap
  samples, each composed of $N=9000$ random samples taken from the full random
  data set of over $10^{4}$ elements.}\BibitemShut {Stop}%
\end{thebibliography}

\end{document}